
\input jnl
{\singlespace\rightline  {NSF-ITP-94-47}\rightline  {gr-qc/9411024}}

\font\tifont=cmssbx10 scaled 1728
\font\lhfont=cmssbx10 scaled 1200
\def\title
  {\null\vskip 3pt plus 0.2fill
   \beginlinemode \doublespace \raggedcenter \tifont}
\def\head#1{
  \goodbreak\vskip 0.5truein
  {\immediate\write16{#1}
   \raggedright\noindent{\lhfont\uppercase{#1}}\par}
   \nobreak\vskip 0.25truein\nobreak}
\def\subhead#1{
  \vskip 0.25truein
  {\raggedright\noindent{\lhfont#1} \par}
   \nobreak\vskip 0.25truein\nobreak}
\def\received{\vskip 3pt plus 0.2fill
 \centerline{\sl (Received November 8, 1994)}}

\title
	Gravitational Collapse of Vacuum Gravitational Field Configurations
\author
	Douglas M. Eardley\footnote%
	  {$^{\rm a}$}{Author's electronic address: \tt doug@itp.ucsb.edu\hfil}
\affil
	Institute for Theoretical Physics
	University of California
	Santa Barbara, CA 93106-4030
\received

\abstract{This paper proves a theorem about the existence of an
apparent horizon in general relativity, which applies equally well to
vacuum configurations and matter configurations. The theorem uses the
reciprocal of the surface-to-volume ratio of a region on a space slice
to measure the radius of the region, and uses the minimum value $K_{\rm
min}$ of certain components of the extrinsic curvature to measure the
strengh of the gravitational field in the region.  The theorem proves
that, if the product of the radius times $K_{\rm min}$ is larger than
unity, then an apparent horizon must form, signalling the formation of
a black hole.}

\body
\oneandathirdspace

\def\refto#1{$^{#1}$}
\citeall\refto
\def\Km{K_{\kern-0.1em\rm min}}
\def\dV{{\partial V}}
\def\area{{\rm Area}}
\def\volume{{\rm Volume}}
\def\Tr{{\rm Tr}}
\def\Min{\mathop{\rm Min}\limits}
\def\U#1{D^{#1}\kern-0.2em f}
\def\L#1{D_{#1}\kern-0.1em f}
\def\Dq{Df^2}
\def\diag{{\rm diag}}
\def\1#1{{1\over#1}}
\def\2#1{{2\over#1}}
\def\3#1{{3\over#1}}
\def\4#1{{4\over#1}}
\def\Eq#1{Eq.~\(#1)}			
\def\Eqs#1{Eqs.~\(#1)}			

\head{I. Introduction}

The ``hoop conjecture" is an informal but plausible assertion about
gravitational collapse due to Kip Thorne.  It claims that a collapsing
body of mass $M$ must be smaller than $\sim M$ in all dimensions before
a black hole can form to enclose it.  This notion led to a series of
important papers in mathematical relativity, starting with the seminal
work of Schoen and Yau,\refto{SY2} proving various theorems about sufficient
conditions on a gravitationally collapsing body so that an apparent
horizon (thus, an event horizon of a black hole) will form.  Most of this
work has concerned the gravitational collapse of matter, even though
the numerical relativists have become adept at the simulation of
gravitational collapse both for matter and for pulses of pure
gravitational radiation.\refto{AE}  A notable result due to Beig
and O'Murchadha\refto{OM} gives sufficient conditions on configurations
of pure gravitational radiation for an apparent horizon to form, but
their result is somewhat hard to apply in practice.

The theorem of Schoen and Yau\refto{SY2} deals with gravitational
collapse of a star, giving sufficient conditions on the radius of the star
and on its energy density for an apparent horizon to form.  Their theorem
depends on the minimum value of the energy density of matter, and so
does not apply to vacuum configurations --- it does not apply to a
``geon'' of gravitational waves in which the gravitational field is so
intense as to cause collapse to a black hole, and it does not even
apply to the maximally extended Schwarzschild geometry.  Moreover,
the particular hypotheses of Schoen and Yau's theorem\refto{SY2}
are violated if the star is deformed by introduction of small vacuoles
in its matter, while it seems clear that the result should continue
to hold.

With these motivations, this paper will use some of Schoen and Yau's ideas
and techniques to prove a further theorem about gravitational collapse,
a theorem that applies equally well to vacuum
configurations, such as the geon and maximally extended Schwarzschild.
This theorem uses the reciprocal of the surface-to-volume ratio of
a region on a space slice to measure the radius of the region, and uses
the minimum value $\Km$ of certain components of the extrinsic
curvature to measure the strengh of the gravitational field in the
region.  The extrinsic curvature $K_{ij}$ is essentially the time derivative
of the 3-metric $g_{ij}$ (for notation see MTW\refto{MTW}) and so $\Km$
can be thought of as measuring the rate at which the spatial geometry is
collapsing.  Since this measure of gravitational field strength makes no
reference to the density of matter, the theorem will apply equally well
to collapsing stars and collapsing gravitational wave pulses.  The theorem
proves that, under the hypothesis that product of the radius time
$K_{\rm min}$ is larger than unity, an apparent horizon must form.
This signals the formation of a black hole and also entails collapse
to a spacetime singularity.  A simple physical interpretation of this
hypothesis is that ``the region is collapsing faster than the speed of
light".

In the sign conventions followed here, a positive sign for a component
of $K_{ij}$ signifies collapse.  Obvious, time-reversed results dual to
those presented here can be derived for superluminal expansion near
the big bang, but these dual results will not be stated or proved in
this paper.  An announcement of this theorem was previously made in
the proceedings of the Texas Symposium.\refto{tex92}

\head{II. Statement of Results}

Let $(N,g_{ij},K_{ij})$ be an initial data set for general relativity.
That is, $(N,g_{ij})$ is a complete Riemannian 3-manifold without boundary,
$g_{ij}$ is the metric of $N$, and $K_{ij}$ is a tensor that
can be interpreted as the extrinsic curvature, when $(N,g_{ij})$ is
imbedded in 4-dimensional spacetime.  Moreover the initial value
equations of general relativity, and the dominant energy condition,
will be assumed to hold.

Consider any bounded region $V$ in $N$ with a smooth boundary $\dV$.  Denote
the outward unit normal vector of $\dV$ by $n^i$, denote the mean extrinsic
curvature of $\dV$ in $N$ in the outward direction by $H$, and denote the
2-dimensional trace of $K_{ij}$ in $\dV$ by
$$
	\Tr_n K \equiv (g^{ij}-n^in^j)K_{ij}.		\eqno(defTrn)
$$
Then $\dV$ is called an {\it apparent horizon} if
$$
	H=\pm Tr_n K
$$
on $\dV$.

Let us now state the asymptotic conditions.  Following Schoen and Yau we
assume that either $(N,g_{ij},K_{ij})$ is asymptotically flat with one
spatial infinity, or, more
generally, that there exists in $N$ a bounded region $V_1$ such that on
$\dV_1$, $K_{ij}$ obeys
$$
	H_1 > \left|Tr_n K\right|.			\eqno(asymp)
$$
In the case of asymptotic flatness, $V_1$ can simply be taken as a large
ball.  Roughly speaking \Eq{asymp} means that there is a large surface
on which gravity is weak.

Now contemplate a region $V \subset V_1$.  In $V$, gravity will be assumed
strong in a precise sense to be explained below.  The goal is to establish
that $N$ is undergoing gravitational collapse to a singularity, and
this will be accomplished by proving that $N$  must contain an apparent
horizon.   It then will follow by the singularity theorems of Penrose
and Hawking that a singularity must occur in spacetime evolution
of the initial data.

Two main quantities enter into the precise definition of ``strong gravity
in the region $V$", namely the size of the region, and the strength of
the gravitational field.
The size of the $V$ region is measured here simply by its surface-to-volume
ratio,
$\area(\dV)/\volume(V)$.
The magnitude of the gravitational field is measured by a quantity $\Km$
which is a certain lower bound on components of the extrinsic curvature
$K_{ij}$.  The precise definition of $\Km$ is
$$
\Km \equiv
  \Min_V \Min_{\{n^i|n^2=1\}} (g^{ij}-n^i n^j)K_{ij}
$$
where $n^i$ is any spatial vector of unit norm.
That is, $\Km$ is defined as follows.  At each point $p\in M$ find the
minimum value of $\Tr_n K$ where $n^i$ is a unit normal to any 2-surface
through $p$, see \Eq{defTrn};  and finally, minimize over all points
$p\in M$.

A simple formula may be given for $\Km$. At each point $p\in M$ there
is an orthonormal basis that diagonalizes $K_{ij}$,
$$
	K_{ij} = \diag(k_1,k_2,k_3), k_1 \le k_2 \le k_3
$$
with the diagonal eigenvalues $(k_1,k_2,k_3)$ ordered as shown; moreover
these eingenvalues are unique at $p$ (even though the basis at $p$ will not
be unique if two of the eigenvalues are equal).  Then $\Km$ is easily seen
to be equal to the sum of smallest two of the three eigenvalues
$(k_1,k_2,k_3)$, minimized over all $p\in M$:
$$
	\Km = \Min_V (k_1+k_2).			\eqno(K12)
$$

Then the result is:
\proclaim Theorem 1.
Let $(N,g_{ij},K_{ij})$ be an initial data set for general relativity.
If $\Km > \area(\dV)/\volume(V)$ for some bounded region $V$ as above,
then an apparent horizon exists within $N$.

\head{III. Proof}

Proof is by contradiction:  We begin by assuming that no apparent
horizon exists in $N$.

The argument is based on Jang's equation, a certain nonlinear partial
differential equation for which Schoen and Yau have given the general
existence theory.  An apparent horizon obstructs the solution of
Jang's equation, and moreover is the {\it only} obstruction under
suitable conditions.\refto{SY1,SY2}  The proof thereby proceeds
by showing that the hypothesis $\Km > \area(\dV)/\volume(V)$ is
inconsistent with the existence of any solution $f$ of Jang's equation.

Jang's equation for a scalar function $f$ on $N$ is
$$
 (1+\Dq)^{-1/2}\left(g^{ij}-{\U i \U j \over1+\Dq}\right)D_i\L j =
	\left(g^{ij}-{\U i \U j\over1+\Dq}\right)K_{ij}.	\eqno(Jang)
$$
Here $D_i$ is the covariant derivative belonging to $g_{ij}$, and
$\Dq \equiv \L i\U i$.  It can also be written in the equivalent form
$$
D^i\left((1+\Dq)^{-1/2}\L i\right) =
	\left(g^{ij}-{\U i \U j\over1+\Dq}\right)K_{ij}.	\eqno(Jang2)
$$

Let us assume that no apparent horizon exists in $V$.  Then by a result of
Schoen and Yau \refto{SY1,SY2} a smooth solution $f$ to Jang's equation exists
in $V$.

Now define the vector field $h^i$ as a renormalized gradient of $f$ by
$$
	h^i \equiv (1+\Dq)^{-1/2}\U i;				\eqno(defh)
$$
note that $h^i$ is of length less than unity,
$$\eqalignno{
	0	&\le h^2 < 1					&(boundh)\cr
\noalign{\noindent where}
	 h^2	&\equiv h^ih_i\cr
}$$
as long as $f$ is smooth and finite.  Jang's equation, \Eq{Jang2}, can now
be written from \Eq{defh} in the equivalent form
$$
	D^ih_i = \left(g^{ij}-h^ih^j\right)K_{ij}.		\eqno(Jang3)
$$
Integrate \Eq{Jang3}over the region $V$ and use Gauss's theorem to yield
$$
  \int_\dV h^i d^2S_i = \int_V \left(g^{ij}-h^ih^j\right)K_{ij}. \eqno(gauss)
$$
By the Cauchy-Schwartz inequality the left hand side of \Eq{gauss} is bounded
by
$$\eqalignno{
  \int_\dV h^i d^2S_i	&\le	\left(\int_\dV d^2S \right)^\12
				\left(\int_\dV h^i h_i d^2S \right)^\12\cr
			&\le	\area(\dV).		&(boundleft)\cr
}$$

Furthermore, the integrand on the right hand side of \Eq{gauss} is bounded
below by $\Km$ as follows.  First note that $\Tr K$ is bounded below by $\Km$.
$$\eqalignno{
  \Tr K	&=	g^{ij}K_{ij}\cr
	&=	\12\sum_{a=1}^3 \left(g^{ij}-e_a^i e_a^j\right)K_{ij}\cr
	&\ge	\32\Km\cr
	&\ge	\Km;\cr
}$$
here $\{e_a^i\}$ $(a=1,2,3)$ is an arbitrarily chosen orthonormal triad;
and in the last line $\Km>0$, by the hypotheses of Theorem 1, has
been used.  Then for the right hand side of \Eq{gauss}
$$\eqalignno{
\left(g^{ij}-h^ih^j\right)K_{ij}
	&=	(1-h^2)\Tr K +h^2(g^{ij}-h^ih^j/h^2)K_{ij}\cr
	&\ge	(1-h^2)\Km +h^2\Km\cr
	&=	\Km,				&(boundright)\cr
}$$
where \Eq{boundh} has been used.
\Eqs{gauss,boundleft,boundright} now imply
$$\eqalignno{
	\area(\dV)		&\ge	\int_V\Km = \volume(V)\Km\cr
\noalign{\noindent or}
	\area(\dV)/\volume(V)	&\ge \Km.\cr
\noalign{\noindent But by the hypotheses of Theorem 1, this is impossible,
as the reverse inequality}
	\area(\dV)/\volume(V)	&< \Km.\cr
}$$
holds.  This contradicts the assumption that no apparent horizon exists
in $N$, and concludes the proof of Theorem 1.

\head{IV. Examples}

\subhead{A. Kasner Universe}

We begin with an example of the local computations of radius and $\Km$.
The Kasner universe is the vacuum Bianchi type I homogeneous anisotropic
solution of the Einstein equations,
$$
	ds^2 = -dt^2 + \sum_{i=1}^3 t^{2p_i}\left(dx^i\right)^2
$$
where the $\{p_i\}$ $(i=1,2,3)$ are three real constants obeying
$$
	p_i\le p_2\le p_3,\qquad \sum_{i=1}^3 p_i = 1 = \sum_{i=1}^3 p_i^2.
$$
Possible values of the $\{p_i\}$  form a one-parameter continuous set,
including $\{0,0,1\}$ and $\{-\13,\23,\23\}$.

In order to model gravitational collapse we shall take $-\infty<t<0$
so that the curvature singularity at $t\rightarrow0$ lies in the future
of all observers in the spacetime, and we shall fix a value of $t<0$
to define a slice $N$.  The extrinsic curvature is
$$
	K_{ij} = \1t\diag(p_1,p_2,p_3)			\eqno(KKas)
$$
and immediately by \Eq{K12}
$$
	\Km = \1t(p_1+p_2).				\eqno(KmKas)
$$
It is easy to see that $\Km>0$ for all permissible $\{p_i\}$, except
for the case (0,0,1) wherein the spacetime is everywhere locally flat.
Now each space slice $t=\const$ is flat; for the region $V$ take a
sphere of some proper radius $R$ in this flat slice.  The
surface-to-volume ratio of $V$ is
$$
	\area(\dV)/\volume(V) = 4\pi R^2/\43\pi R^3 = 3/R.	\eqno(AKas)
$$
If we choose
$$
	R > 3/\Km = 3t/(p_1+p_2)
$$
the local condition for existence of a trapped surface is satisfied.
However, the Kasner universe is not asymptotically flat, so Theorem 1
does not apply directly.  Therefore, turn immediately to
an asymptotically flat example which resembles the Kasner universe
near the singularity, namely the Schwarzwschild black hole.

\subhead{B. Schwarzschild Black Hole}

Consider the spherically symmetric vacuum black hole\refto{MTW} interior
$r<2M$ in
$$
 ds^2 = -\left({2M\over r}-1\right)^{-1}dr^2 +
	 \left({2M\over r}-1\right)dt^2 + r^2(d\theta^2+\sin^2\theta d\phi^2)
$$
and consider a slice $N$, topologically $S_2\times R^1$ defined by
$r=\const$, $-\infty<t<\infty$.  In order to apply Theorem 1, the slice
$N$ must be deformed so that it is asymptotically flat: This deformation can
be done by pushing the ``ends" of the slice out of the black hole and off to
spatial infinity, outside a large region $-T<t<T$ where $T \gg M$ is
some constant.

Inside the black hole, $r$ is a timelike coordinate, however it decreases
toward the future.  Therefore it is convenient to take $r'=-r$ as the
timelike coordinate in defining $K_{ij}$.  Then in the coordinate basis
$\{t,\theta,\phi\}$, the extrinsic curvature is calculated as
$$
 K_{ij} = \left({2M\over r}-1\right)^\12\diag(-{M\over r^2},r,r\sin^2\theta)
$$
and the eigenvalues $(k_1,k_2,k_3)$ in the corresponding orthonormal
basis are
$$
	     \left(-\left({2M\over r}-1\right)^{-\12}{M\over r^2},
		    \left({2M\over r}-1\right)^\12{1\over r},
		    \left({2M\over r}-1\right)^\12{1\over r}\right) \eqno(KS)
$$
and using \Eqs{K12,KS}
$$
  \Km = r^{-1}\left({2M\over r}-1\right)^{-\12}
				\left({M\over r}-1\right).	\eqno(KmS)
$$
Thus $\Km>0$ for $r<M$, and as we approach the singularity
$$
 \Km \approx \left({M\over2r^3}\right)^\12\hbox{\quad as\quad}r\rightarrow0.
								\eqno(KmS0)
$$
This result for $\Km$ is approximately equal to that of the Kasner
universe above, \Eq{KmKas} with $\{p_1,p_2,p_3\}=\{-\13,\23,\23\}$,
and with
$$
	t_{\rm Kasner}\approx\sqrt{2r^3\over9M}.
$$

For the region $V$ in the slice $r=\const\ll M$, take a small
coordinate ball $V_\delta$ of some proper radius $\delta$
centered at $t=t_0$,$\theta=0$,
$$
  V_\delta = \left\{\quad(t,\theta,\phi)\quad\mathop\Bigg|\quad\left({2M
\over r}-1\right)(t-t_0)^2 + r^2\theta^2<\delta^2,0\le\phi<2\pi\quad\right\}.
$$
For
$$
\delta/r\ll 1,							\eqno(dsmall)
$$
$V_\delta$ is approximately a spherical ball in an approximately flat region,
and its surface-to-volume ratio is (compare \Eq{AKas})
$$
  \area(\dV_\delta)/\volume(V_\delta)	\approx3/\delta.	\eqno(AS0)
$$
Comparing \Eqs{KmS0, AS0}, Theorem 1 applies for
$$\eqalignno{
	\left({M\over2r^3}\right)^\12 &\gtwid {3\over\delta}\cr
\noalign{\noindent or, also taking the restriction, \Eq{dsmall}, into account}
  r\gg \delta	&\gtwid \left({18r^3\over M}\right)^\12.	&(dlim)\cr
}$$
It is now clear from \Eq{dlim} that, for any sufficiently small $r\ll M$,
a choice of $\delta$ exists so that Theorem 1 applies.  Thus the slice
$N$ must contain a trapped surface.

\subhead{C.  Gravitational Collapse to a Schwarzschild Black Hole}

Consider a Schwarzschild black hole formed by gravitational collapse of
a star or gravitational wave pulse.  We assume that the hole relaxes to
the Schwarzschild state at large time, including near the singularity;
this will always be true, for instance, in strict spherical symmetry.
In this case, the slice $N$ has only one end (for large $t$), which is
pushed off to spatial infinity; at small $t$  the slice penetrates the
collapsing region.  The slice is therefore topologically $R^3$.

The computations above still apply and the same conclusion holds: For
suitable $\delta$, the slice $N$ must contain a trapped surface.  This
result is interesting in light of a result of Wald and Iyer.\refto{WI}
They point out that, in a Schwarzschild black hole formed from
spherical gravitational collapse, there may exist Cauchy slices $N$
that come arbitrarily near to the singularity, at late times in the
interior of the vacuum black hole, yet do not contain any trapped
surfaces.  Theorem 1 can be used to bound this behavior:  Any such
slice $N$ that contains a small coordinate ball $V_\delta$ with
$\delta$ large enough, \Eq{dlim}, cannot fail to to contain a trapped
surface.

\subhead{D. Possible Generalizations}

One would like to have a generalization of Theorem 1 which omits
or weakens the asymptotic conditions, \Eq{asymp} and above.  Simple
examples, such as the Kasner solution above, do in fact admit trapped
surfaces near the singularity, despite the fact that they do not
obey our asymptotic condition.  This encourages the conjecture that
these conditions can be dispensed with somehow.

One might also wish for more information on the location of trapped
surfaces whose existence is guaranteed by Theorem 1.  In known examples,
trapped surfaces intersect the region $V$, and we conjecture this is
always true.  However, simple counterexamples also show that $V$ may not
actually contain any trapped surfaces.  For instance, take the
Schwarzschild example above, but modify the region $V=V_\delta$, as
follows.  Deform $V_\delta$ by pushing a narrow finger of its surface
a long ways into the interior.  This deformation will not change the
surface-to-volume ratio very much.  However, the finger can clearly
be made to intersect any trapped surfaces that lie entirely within
$V_\delta$, providing the required counterexample.

\subhead{E. Possible Applications to Numerical Relativity}

Much of numerical relativity is carried out with the maximal slicing
condition,
$$
	\Tr K = 0 = k_1+k_2+k_3.
$$
Theorem 1 cannot apply in this case, because from \Eq{K12}
$$
	\Km = k_1 + k_2 = -k_3 \le 0
$$
while we must have $\Km>0$ for Theorem 1 to be of any use.  However,
some interesting numerical work has been done with constant-mean-curvature
slicing,
$$
	\Tr K = \const
$$
or with $\Tr K$ as a time function, or with a more general related choice
of time function,\refto{Ove} and Theorem 1 may well apply in these cases.

\head{Acknowledgements}

This research was supported in part by the National Science Foundation
under Grant Nos.~PHY89-04035 and PHY90-08502.  I have been greatly
helped by discussions with various people at ITP, at USCB, and at
the Aspen Center for Physics.

\references
\oneandathirdspace

\refis{SY1} R. Schoen \& S.-T. Yau, \cmp 79, 231-260, 1981.

\refis{SY2} R. Schoen \& S.-T. Yau, \cmp 90, 575-579, 1983.

\refis{OM}  R. Beig \& N. O'Murchadha, \prl 66, 2421-2424, 1991.

\refis{AE}  A.M. Abrahams and C.R. Evans, \prd 46, 4117-4121, 1992.

\refis{tex92} D.M. Eardley, \journal Ann. N.Y. Acad. Sci., 688, 408-417, 1993.

\refis{MTW} C.W. Misner, K.S. Thorne and J.A. Wheeler,
{\it Gravitation} (W.H. Freeman.  San Francisco, 1973).

\refis{WI} R.M. Wald and V. Iyer, \prd 44, 3719-3722, 1991.

\refis{Ove} R. Ove, \jmp 31, 2688-2693, 1990.

\endreferences

\endit